\documentclass[a4paper]{article}
\usepackage{epsfig}
\usepackage{graphicx}  
\usepackage{color}     
\usepackage{lscape}
\usepackage{color}
\date{May 20, 2012}
\begin{document}

\title{\Large \bf Modified Chaplygin Gas Cosmology }
{\small
\author{H. B. Benaoum \\
Prince Mohammad Bin Fahd University, Al-Khobar 31952, Saudi Arabia  \\
Email: hbenaoum@pmu.edu.sa}
}

\maketitle
\begin{center}
\small{\bf Abstract}\\[3mm]
\end{center}
Modified Chaplygin gas as an exotic fluid has been introduced in \cite{benaoum}.
Essential features of the modified Chaplygin gas as a cosmological model are discussed. Observational constraints on the parameters of the model have been 
included. The relationship between the modified Chaplygin gas and a homogeneous minimally coupled scalar field are reevaluated 
by constructing its self-interacting potential. In addition, we study the role of the tachyonic field in the modified Chaplygin gas 
cosmological model and the mapping between scalar field and tachyonic field is also considered.
\\\\

\begin{minipage}[h]{14.0cm}
\end{minipage}
\vskip 0.3cm \hrule \vskip 0.5cm

\newpage
\section{Introduction} 
Astronomical and cosmological observations, such as type Ia supernovae (SNe~Ia) \cite{per1}-\cite{riess2}, large scale redshift surveys structure (LSS) \cite{bachall,tegmark}, the cosmic microwave background (CMB) \cite{miller,bennet} 
and  Wilkinson Microwave Anisotropy Probe (WMAP) \cite{briddle,spergel}, indicate that the observable universe 
experiences an accelerated expansion. 
These observations suggest also that the universe is nearly flat and dominated by a non-baryonic substratum. 
The source of this acceleration is usually attributed to an exotic type of fluid with negative pressure called 
commonly dark energy. \\
Various kind of dark energy models have been proposed such as cosmological constant \cite{peebles1}, quintessence \cite{ratra}-\cite{sami}, k-essence \cite{picon}-\cite{scherrer}, tachyon \cite{sen1}-\cite{gibbons}, 
phantom \cite{caldwell}-\cite{cline}, Chaplygin gas \cite{pasquier}, 
quintom \cite{feng}, holographic dark energy \cite{horova} and extra dimensions \cite{dvali}. 
The nature of the dark sector of the universe (i.e. dark energy and dark matter) remains a mystery. An 
economical and attractive idea to unify the dark sector of the universe is to consider it as a single component that acts 
as both dark energy and dark matter. One way to achieve the unification of dark energy and dark matter is by using 
the so-called Chaplygin gas. The pure Chaplygin gas or generalized Chaplygin gas is a perfect fluid which behaves like 
a pressureless fluid at an early stage and a cosmological constant at a later stage. \\
Pure Chaplygin gas with an exotic equation of state is characterized by a negative pressure \cite{pasquier}, 
\begin{eqnarray}
p = - \frac{B}{\rho}
\end{eqnarray}
where $p$ is the pressure, $\rho$ the energy density and $B$ a positive parameter. \\
The pure Chaplygin gas has been extended to the so-called generalized Chaplygin gas with the following equation 
of state \cite{bento}, 
\begin{eqnarray}
p = - \frac{B}{\rho^{\alpha}}  ~~~~~\mbox{where~~~$0 \le \alpha \le 1$} .
\end{eqnarray}
It is clear that the pure Chaplygin gas is recovered for the case $\alpha = 1$. \\
The interesting feature of the Chaplygin gas is its connection to the string theory. It can be obtained from the 
Nambu-Goto action for a D-brane moving in a (D+2)-dimensional spacetime in the light cone parametrization \cite{souza}-\cite{ogawa}. \\
The outline of this paper is as follows. In the next section, we study the cosmological model of the modified Chaplygin 
gas which was introduced by the author in \cite{benaoum}. We show that the modified Chaplygin gas model 
interpolates between an epoch with a soft equation of state and a de Sitter phase. 
In section 3, the relationships between the modified Chaplygin gas and a cosmological scalar field are reevaluated. 
In section 4, a tachyonic field is considered as a candidate for the modified Chaplygin gas model. 
The correspondence between minimally coupled scalar field and the tachyon field is also investigated. 

\section{FRW Cosmology from the Modified Chaplygin gas Model}
Within the framework of Friedmann-Robertson-Walker (FRW) cosmology, a model called modified Chaplygin gas has been proposed by the author \cite{benaoum}. 
This model includes an initial phase of radiation and it is based on the following equation of state : 
\begin{eqnarray}
p & = & A \rho - \frac{B}{\rho^{\alpha}}  
\end{eqnarray}
where  $A, B$ and $\alpha$ are constant parameters. \\  
When $B = 0$, we recover the equation of state of perfect fluid, i.e. $p = A \rho$. For $A=0$, it reduces to the 
generalized Chaplygin gas. \\
In equation (3), the two terms start to be of the same order when the pressure 
vanishes ( i.e. $p = 0$ ). In this case, the fluid has pressureless 
density $\rho_0$, corresponding to some scale factor $a_0$ , 
\begin{eqnarray}
\rho_0 & = & \rho^{\alpha+1} ( a_0 )~=~ \frac{B}{A} ~~.
\end{eqnarray}
The metric of $D$-dimensional FRW spacetime is :
\begin{eqnarray}
d s^2 & = & - d t^2 + a{^2}(t)~ d \Omega_k{^2}
\end{eqnarray} 
where $a (t)$ is the scale factor and $d \Omega_k{^2}$ is the metric of the maximally symmetric $( D - 1)$--
space with curvature $k = 0,\pm 1$ . \\  

The Friedmann equation which governs the evolution of the scale factor is given by : 
\begin{eqnarray}
H^2 & = & \left( \frac{\dot{a}}{a} \right)^2~ =~ 
\frac{2~\rho}{(D-1) (D-2)} - \frac{k}{a^2} 
\end{eqnarray}
where $H$ is the Hubble parameter. \\

In the FRW framework, a fluid with an energy density $\rho$ and a pressure $p$ must satisfy the conservation law : 
\begin{eqnarray} 
\dot{\rho} + (D-1) H ( \rho + p ) & = & 0 
\end{eqnarray}   

The latter two equations imply that :
\begin{eqnarray}
\frac{\ddot{a}}{a} & = & \dot{H} + H^2~ =~ -~
\frac{(D-1)~p + (D-3)~\rho}{(D-1)(D-2)} ~~~~.
\end{eqnarray}

By defining $W = a^{(D - 1) (A + 1)}$ and a rescaled density 
$\bar{\rho} = \rho W$, the equation (7) becomes, 
\begin{eqnarray}
\dot{\bar{\rho}} - \frac{B}{A + 1}~\frac{W^{\alpha} \dot{W}}{\bar{\rho}^{\alpha}} & = & 0 ~~.
\end{eqnarray} 
This equation can be easily integrated leading to :  
\begin{eqnarray}
\frac{\bar{\rho}^{\alpha+1}}{\alpha + 1} & = & \frac{B}{A+1} \frac{W^{\alpha+1}}{\alpha+1}~ 
+~ \frac{C}{\alpha+1}  
\end{eqnarray}
where $C$ is an integration constant . \\
The density will be : 
\begin{eqnarray}
\rho & = & \left( \frac{B}{A+1} 
+ \frac{C}{W^{\alpha+1}} \right)^{\frac{1}{\alpha+1}} ~~~.
\end{eqnarray} 
The constant of integration $C$ can be expressed in terms of the 
cosmological scale $a_0$,~( i.e. $W_0 = a_0^{(D -1 )(A+1)}$~) 
where the fluid has a vanishing pressure :  
\begin{eqnarray} 
C & = & \frac{B}{A+1}~ \frac{W_0^{\alpha+1}}{A} 
\end{eqnarray} 
The energy density $\rho$ will be : 
\begin{eqnarray}
\rho & = & \left( \frac{B}{A+1} \right)^{\frac{1}{\alpha+1}}~ \left( 1 + 
\frac{1}{A~W_r^{\alpha+1}}  \right)^{\frac{1}{\alpha+1}}  
\end{eqnarray} 
where $W_r = W/W_0$. \\

For large scale factor $a$, i.e. $W_r >> 1$, we have : 
\begin{eqnarray} 
\rho & \simeq & \left( {\frac{B}{A+1}} \right)^{\frac{1}{\alpha+1}}  \nonumber \\
p & \simeq & - \left({\frac{B}{A+1}} \right)^{\frac{1}{\alpha+1}} = - \rho 
\end{eqnarray} 
which correspond to an empty universe with a cosmological constant 
$\left( {\frac{B}{A+1}} \right)^{\frac{1}{\alpha+1}}$ (i.e. a de Sitter space ). \\
Also for small scale factor $a$, i.e.  $W_r << 1$, we have :  
\begin{eqnarray} 
\rho & \simeq & \left( \frac{B}{A (A+1)} \right)^{\frac{1}{\alpha+1}} ~\frac{1}{W_r} \nonumber \\
p & \simeq & A \left( \frac{A}{B (B+1)} \right)^{\frac{1}{\alpha+1}} ~\frac{1}{W_r} = A \rho 
\end{eqnarray} 
which correspond to universe dominated by an equation of state 
$p = A \rho$. This shows that this model interpolates between a universe 
dominated by matter phase with equation of state $p = A \rho$ and 
a de Sitter phase $p \simeq - \rho$. \\ 
Moreover, expanding equations (13) and (3) to the subleading terms at large 
cosmological constant, we obtain the following expressions for the energy 
and the pressure : 
\begin{eqnarray}
\rho & \simeq & \left( {\frac{B}{A+1}} \right)^{\frac{1}{\alpha+1}}~
\left( 1 + \frac{1}{ ( \alpha+1) A W_r^{\alpha+1}}  \right) \nonumber \\
p & \simeq & \left( {\frac{B}{A+1}} \right)^{\frac{1}{\alpha+1}}~ 
\left( - 1 + \frac{A + \alpha (A + 1)}{( \alpha+1) A W_r^{\alpha+1}}  \right)  ~~.
\end{eqnarray} 
These correspond to the mixture of a cosmological constant 
$\left( \frac{B}{A+1} \right)^{\frac{1}{\alpha+1}}$ and a type of matter 
described by an equation of state : 
\begin{eqnarray}
p & = & \left( \alpha + A (\alpha + 1) \right)~ \rho ~~~~.
\end{eqnarray}  
The equation of state parameter takes the form :
\begin{eqnarray}
\omega & = & \frac{p}{\rho}~=~ -~ 
\frac{1 - \frac{1}{W_r^{\alpha+1}}}{1 + \frac{1}{A W_r^{\alpha+1}}} 
\end{eqnarray}
which ranges over $-1 < \omega < A$ , depending on the cosmological scale $a$, 
\begin{eqnarray}
\omega &= & \left\{ \begin{array}{cl} 
-1 & \mbox{for}~~W_r >> 1 \\ 
0  & \mbox{for}~~W_r = 1 \\
A  & \mbox{for}~~W_r << 1 ~~~~~. \end{array} 
\right.
\end{eqnarray}
The speed of sound $c_s$ is defined as, 
\begin{eqnarray} 
c_s^2 & = & \frac{\partial p}{\partial \rho}~=~ \frac{\dot{p}}{\dot{\rho}} ~~.
\end{eqnarray} 
Now by computing $\dot{\omega}$, we obtain : 
\begin{eqnarray} 
\dot{\omega} & = & \left( c_s^2 - \omega \right)~ \frac{\dot{\rho}}{\rho}  
\end{eqnarray} 
which for the modified Chaplygin gas gives the following expression for the speed of sound :   
\begin{eqnarray} 
c_s^2 & = & \omega + \rho \frac{d \omega}{d \rho} ~ = ~ 
\frac{\alpha + (\alpha +1) A  + \frac{1}{W_r^{\alpha+1}}}{1 + \frac{1}{A W_r^{\alpha+1}}} 
\end{eqnarray} 
implying that $c_s^2$ is always positive and hence there is no concern about imaginary speed of sound. \\ 
Moreover it has the following asymptotic limit, 
\begin{eqnarray}   
c_s^2 & = & \left\{ \begin{array}{cl}   
\alpha + (\alpha+1) A & \mbox{for}~~W_r >> 1 \\ 
(\alpha+1) A  & \mbox{for}~~W_r = 1 \\
A  & \mbox{for}~~W_r << 1 ~~~~. \end{array} \right.
\end{eqnarray}  
The speed of sound never exceeds that of light for smaller scale $a$ or scale of the order of $a_0$ 
where the pressure vanishes, provided that $A < \frac{1}{\alpha + 1}$ and will exceed it for large 
scale compared to $a_0$. \\  

The constraints from the astrophysical and cosmological observables on the modified Chaplygin gas has been studied by many authors \cite{lu1}-\cite{fabris}. 
The permissible values of the parameters $A, B$ and $\alpha$ have been explored from the observed data. \\
Using data from different observations, namely, observational Hubble Data (OHD), baryon acoustic oscillation (BAO) and CMB shift parameters data, the allowed values for some of the parameters of the modified Chaplygin gas have been extracted. In the light of the 182 Gold SNe Ia, the 3-year WMAP and the SDSS data, the best fits correspond to $B = -0.085$ and $\alpha = 1.724$ \cite{lu1}.
This result was obtained by decomposing the modified Chaplygin gas into two components, i.e. dark matter and dark energy component. \\
However, by using Markov Chain Monte Carlo Method with the observational data from the SN Ia Union 2, OHD, cluster X-rays gas mass fraction (CBF), BAO and CMB data, the 
best fits of the modified Chaplygin gas parameters give $B = 0.00189$ and $\alpha = 0.1079$ \cite{lu2}. \\
Moreover, a perturbative analysis of the modified Chaplygin gas shows that the power spectrum observational data restricts the value of 
$|B| < 10^{-6}$ \cite{fabris} such that the modified Chaplygin gas is disfavored. 

\section{Modified Chaplygin Gas as a Scalar Field}
Following \cite{barrow1}-\cite{nakamura}, we describe the modified Chaplygin gas cosmological model by introducing a 
scalar fields having a self-interacting potential $U(\varphi)$ with the Lagrangian :
\begin{eqnarray}
{\cal L}_{\varphi} & = & \frac{\dot{\varphi}^2}{2} - U(\varphi) ~~. 
\end{eqnarray}
Both the energy density and the pressure of the modified Chaplygin gas can be related to the scalar $\varphi$ through the following transformation equations : 
\begin{eqnarray}
\rho_{\varphi} & = & \frac{\dot{\varphi}^2}{2} + U(\varphi) =~ \rho  \nonumber \\
p_{\varphi} & = & \frac{\dot{\varphi}^2}{2} - U(\varphi) =~ A \rho - \frac{B}{\rho^\alpha} ~~~~.
\end{eqnarray} 
The kinetic energy of the scalar field $\varphi$ and its corresponding potential $U(\varphi)$ are :
\begin{eqnarray}
\dot{\varphi}^2 & = & ( 1 + \omega_{\varphi} )~ \rho_{\varphi}  \nonumber \\
U(\varphi) & = & \frac{1}{2}~( 1 - \omega_{\varphi} )~\rho_{\varphi} 
\end{eqnarray} 
where $\omega_{\varphi} = p_{\varphi}/\rho_{\varphi}$ ~. \\

Now since $\dot{\varphi} = \varphi' \dot{W_r}$ where the prime denotes derivation 
with respect to $W_r$ and $\dot{W_r} = (D -1) (A + 1) H W_r$, we get : 
\begin{eqnarray}
\varphi'{^2} & = & \frac{D -2}{2 (D -1) (A+1)^2}~\frac{1 + \omega_{\varphi}}{W_r^2} ~~.
\end{eqnarray} 
Here we have used equation (6) for the Hubble constant and guided by the cosmic 
microwave background CMB data which is strongly consistent with a flat 
universe, and restricted ourselves to the flat case $k = 0$. \\ 
By using equation (18), we get : 
\begin{eqnarray}
\varphi' & = & \sqrt{\frac{D -2}{2 (D -1) A (A +1)}}~ 
\frac{1}{W_r^\frac{\alpha+3}{2} \sqrt{1 + \frac{1}{A W_r^{\alpha+1}}}} \nonumber \\
U \left( \varphi \right) & = & \frac{1}{2} 
\left(\frac{B}{A+1} \right)^{\frac{1}{2}}~ \frac{2 + \frac{1 -A}{A W_r^{\alpha+1}}}{\sqrt{1 + \frac{1}{A W_r^{\alpha+1}}} } ~~.
\end{eqnarray}  
The first equation can be integrated easily which gives : 
\begin{eqnarray}
A W_r^{\alpha+1} & = & \frac{1}{ \sinh^2 \left( \varpi (\alpha+1)  \Delta \varphi \right)} 
\end{eqnarray}
where $\varpi = \sqrt{\frac{(D-1)(A+1)}{2 (D-2)}}$ and $\Delta \varphi = \varphi - \varphi_0$. \\
We note that for larger scales, the scalar field asymptotically approaches to the constant field $\varphi_0$ and becomes infinite 
(i.e. $\phi \rightarrow \ + \infty$ ) for small scales. \\
Next, by substituting the latter expression in equation (25), we can write all our physical quantities 
$\rho_{\varphi}, p_{\varphi}$ and $\omega_{\varphi}$ in terms of the scalar field $\varphi$ as, 
\begin{eqnarray}
\rho_{\varphi} & = & \left( \frac{B}{A + 1} \right)^{\frac{1}{\alpha+1}}~
\cosh^{\frac{2}{\alpha+1}} \left( \varpi (\alpha+1) \Delta \varphi   \right) \nonumber \\
p_{\varphi} & = & \left( \frac{B}{A + 1} \right)^{\frac{1}{\alpha+1}}~
\left[ A~\cosh^{\frac{2}{\alpha+1}} \left( \varpi (\alpha+1)  \Delta \varphi \right) - 
\frac{A+1}{\cosh^{\frac{2 \alpha}{\alpha+1}} \left( \varpi (\alpha+1) \Delta \varphi \right)} \right] \nonumber \\
\omega_{\varphi} & = & -~\frac{1 - A~\sinh^2 \left( \varpi (\alpha+1) \Delta \varphi \right)}{\cosh^2 \left(\varpi (\alpha+1) \Delta \varphi \right)} ~~.
\end{eqnarray}
Notice that these physical quantities do not depend on the intermediate 
constant $W_0$ (i.e. constant of integration $C$ ). \\
Finally, we get the following potential which has a simple form :
\begin{eqnarray}
U ( \varphi) & = & \frac{1}{2} \left( \frac{B}{A + 1} \right)^{\frac{1}{\alpha+1}}~ 
\left[ \frac{1 + A}{\cosh^{\frac{2 \alpha}{\alpha+1}} \left( \varpi (\alpha+1) \Delta \varphi \right)} + 
(1  - A) \cosh^{\frac{2}{\alpha+1}} \left( \varpi (\alpha+1) \Delta \varphi \right) \right] ~~~.
\end{eqnarray} 
\section{Modified Chaplygin Gas as Tachyonic Field }  
The importance of tachyon in cosmology is inspired by string theory \cite{sen1}-\cite{sen2}. 
The action of the homogeneous tachyon condensate of string theory in a gravitational background is given by : 
\begin{eqnarray} 
S & = & \int d^D x \sqrt{- g}~\left( R + {\cal L}_{tachyon} \right) 
\end{eqnarray} 
where $R$ is the scalar curvature. For a tachyonic field $T$ with tachyonic potential $V(T)$, the relativistic Lagrangian can be expresses as :
\begin{eqnarray}
{\cal L}_{tachyon} & = & - V(T)~\sqrt{ 1 + g^{\mu \nu} ~~.
\nabla_{\mu} T \nabla_{\nu} T} ~~~~. 
\end{eqnarray}
The corresponding energy-momentum tensor for the tachyonic field is : 
\begin{eqnarray}
T_{\mu \nu} & = & - \frac{2 \delta S}{\sqrt{- g} \delta g^{\mu \nu}} 
~= ~p_T~ g_{\mu \nu} + \left( p_T + \rho_T \right)~ u_{\mu} u_{\nu} 
\end{eqnarray}
where the velocity $u_{\mu}$ is : 
\begin{eqnarray}
u_{\mu} & = & \frac{- \nabla_{\mu} T}{\sqrt{ - g^{\mu \nu} \nabla_{\mu} T   
\nabla_{\nu} T }}
\end{eqnarray}  
with $u^{\mu} u_{\mu} = - 1$ ~. \\

It follows that the energy density $\rho_T$ and the pressure $p_T$ of the tachyonic field are given by : 
\begin{eqnarray} 
\rho_T & = & \frac{V(T)}{\sqrt{1 - \dot{T}^2}} \nonumber \\
p_T & = & - V(T) \sqrt{ 1 - \dot{T}^2 } ~~~.
\end{eqnarray}
The equation of state parameter is : 
\begin{eqnarray}
\omega_T & = & \frac{p_T}{\rho_T}~=~ - \left(1 -  \dot{T}^2 \right) ~~.
\end{eqnarray} 
The condition of accelerating universe (i.e. $\frac{\ddot{a}}{a} > 0$ ) 
requires that : 
\begin{eqnarray}
\dot{T}^2 < \frac{2}{D - 1 } 
\end{eqnarray}
and the rolling tachyon has an interesting equation of state whose parameter $\omega_T$ interpolates between $-1$ and 
$\frac{-D+3}{D-1}$. \\
The evolution of the tachyonic field is driven by : 
\begin{eqnarray}
\ddot{T} + (1 - \dot{T}^2 ) \left( (D-1) H \dot{T} + \frac{1}{V (T)} \frac{d V(T)}{d T} \right) & = & 0
\end{eqnarray}
with the constraint equation for the Hubble parameter give by : 
\begin{eqnarray}
H^2 & = & \frac{2}{(D-1) (D-2)} \frac{V(T)}{\sqrt{1 - \dot{T}^2}} \nonumber \\
\dot{H} & = & - \frac{1}{D-1} \frac{ \dot{T}^2 ~V(T)}{\sqrt{1 - \dot{T}^2}} ~~~.
\end{eqnarray}
By combining the last two equations, the tachyonic field $T$ and the potential $V (T)$ can be expressed as : 
\begin{eqnarray}
T \left( t \right) & = & \int dt ~\left( - \frac{2}{D-1} \frac{\dot{H}}{H^2} \right)^{1/2} \nonumber \\
V \left( T \right) & = & \frac{(D-1) (D-2)}{2} H^2 \left( 1 +\frac{2}{D-1} \frac{\dot{H}}{H^2} \right)^{1/2} ~~.
\end{eqnarray}
Note that the knowledge of $H$ and $\dot{H}$ ( i.e. cosmological scale factor $a (t)$ ) completely determines the 
tachyonic field $T$ and its corresponding potential $V (T)$. \\

By mapping the pressure $p_{\varphi}$ and the energy density $\rho_{\varphi}$ for 
the scalar field with the corresponding tachyon field $p_T$ and $\rho_T$, we obtain that : 
\begin{eqnarray}
d T & = & \frac{d \varphi}{\sqrt{\rho_{\varphi}}} \nonumber \\
V ( T ) & = & \sqrt{- \omega_{\varphi}} ~\rho_{\varphi} ~~~.
\end{eqnarray} 
Cosmological correspondence between the tachyonic field and a minimally coupled scalar field has been noted by \cite{pad,benaoum}. 
Such correspondence was also investigated by \cite{gorini} where it was demonstrated explicitly that distinct scalar field 
and tachyonic models may give rise to the same cosmological evolution for a particular choice of initial conditions. \\

Now by using equation (30), an exact integration to the first equation in (42) can be performed without any approximation as follows :  
\begin{eqnarray}
T - T_0  & = &  \frac{\cosh^{\frac{\alpha}{\alpha+1}} \left(\varpi (\alpha+1) \Delta \varphi \right)}{\alpha \varpi~\left(\frac{B}{A+1}\right)^{\frac{1}{2 (\alpha+1)}}} ~
{}_2F_1 \!\left[ \begin{array}{cc}
\frac{1}{2},\frac{1}{2} -\frac{1}{2 (\alpha +1)} \\
~~~\frac{3}{2}- \frac{1}{2 (\alpha +1)} \end{array} ;
- \cosh^2 \left(\varpi (\alpha+1) \Delta \varphi \right) \right]
\end{eqnarray}
where ${}_2F_1$ is the hypergeometric function given by : 
\begin{eqnarray}
{}_2F_1 \!\left[ \begin{array}{cc}
a, b \\
~~~c \end{array} ;
x \right] & = & \sum_{k=0}^{[\infty} \frac{ (a)_k (b)_k}{(c)_k}~
\frac{x^k}{k!} ~~.
\end{eqnarray}
with $(a)_k = a \left( a + s \right) \ldots \left( a + (n - 1) s \right)$ as the Pochhammer's symbol. \\
Furthermore, by replacing the first equation in (30), the equation (43) for tachyonic field $T$ can be written in terms of the energy density $\rho_{\varphi}$ as: 
\begin{eqnarray}
T - T_0  & = &  \frac{\rho^{\frac{\alpha}{2}}_{\varphi}}{\alpha \varpi~\left( \frac{B}{A+1} \right)^{\frac{1}{2}}} ~
{}_2F_1 \!\left[ \begin{array}{cc}
\frac{1}{2},\frac{1}{2} -\frac{1}{2 (\alpha +1)} \\
~~~\frac{3}{2}- \frac{1}{2 (\alpha +1)} \end{array} ;
- \left(\frac{A+1}{B} \right) ~\rho^{\alpha+1}_{\varphi} \right] ~~ .
\end{eqnarray}

To find the scalar field $\varphi$ and its potential $U (\varphi)$ in terms of the tachyon field and its potential $V (T)$, one has to perform the inverse transformation : 
\begin{eqnarray}
d \varphi & = & \sqrt{\rho_T} ~d T \nonumber \\
U (\varphi) & = & \frac{1}{2} \left( 1 - \omega_T \right) ~\rho_T ~~~.
\end{eqnarray}

In the slow-rolling approximation, the tachyon field potential $V (T)$ and the scalar field potential $U (\varphi)$ are approximately the same. To see this, we expand 
equation (41) up to first order in $\dot{H}/H^2$ and use equations (6) and (8) to get : 
\begin{eqnarray}
V (T) & \simeq \frac{(D-1) (D-2)}{2} ~\left( H^2 + \frac{1}{D-1} ~\dot{H} \right) = \frac{1}{2} ~\left( 1 - \omega_{\varphi} \right) ~\rho_{\varphi} = U (\varphi ) ~~.
\end{eqnarray}
Now to rewrite the potential $V (T)$ in terms of $T$, we first expand the equation (45) to first order in $\rho_{\varphi}$ :
\begin{eqnarray}
T - T_0 & \simeq & \frac{\rho^{\frac{\alpha}{2}}_{\varphi}}{\alpha \varpi ~\left( \frac{B}{A+1} \right)^{\frac{1}{2}}} 
\end{eqnarray} 
and then substitute it in equation (47) to finally get : 
\begin{eqnarray}
V (T) & \simeq & \frac{c_1}{(T- T_0)^2} + c_2 ~(T- T_0)^2 
\end{eqnarray}
where $c_1$ and $c_2$ are constant parameters depending on $D, \alpha, A$ and $B$. 
\section{Conclusions }
One approach in modern cosmology consists in supposing that dark energy and dark matter are different manifestations of a single entity. Following such an idea, 
this work ( see also \cite{benaoum} ) presents a cosmological model based on the modified Chaplygin gas that acts as a single component. It is shown that the equation 
of state of the modified Chaplygin gas interpolates from matter-dominated era to a cosmological constant dominated era. 
Astronomical and cosmological constraints on the parameters $A, B$ and $\alpha$ of the modified Chaplygin have been investigated and still no basic agreement 
is reached on their values. \\
In addition, from the theoretical point of view the modified Chaplygin gas model is equivalent to that of a scalar field $\varphi$ having 
self-interacting potential $U (\varphi)$. 
Such a description has been explored and its self-interacting potential has been determined. \\
Furthermore, it has been shown that the modified Chaplygin gas can also be described by a tachyonic field $T$ having a potential $V (T)$. Correspondence between scalar field and tachyonic field descriptions has been investigated. The transformations between the scalar field and tachyonic field and between their corresponding potentials have 
been determined. Such a correspondence has been applied to derive the exact expression of the tachyonic field $T$ in terms of the scalar field $\varphi$ for the modified Chaplygin gas. Finally, in the slow-rolling approximation, we have expressed the the potential $V (T)$ in terms of the tachyonic field $T$.  

\end{document}